*Article*

# Resolving the effects of nanoscale membrane curvature on lipid mobility

**Abir Maarouf Kabbani**[1,†], **Xinxin Woodward**[1,†], and **Christopher V. Kelly**[1,*]

Department of Physics and Astronomy, Wayne State University
[†] These authors contributed equally to this work.
* Correspondence: cvkelly@wayne.edu; Tel.: +1-313-577-8471

**Abstract:** The biophysical consequences of nanoscale curvature have been challenging to resolve due to size-dependent membrane behavior and the experimental resolution limits imposed by optical diffraction. Recent advances in nanoengineering and super-resolution techniques have enabled new capabilities for creating and observing curvature. In particular, draping supported lipid bilayers over lithographically patterned substrates provides a model system for endocytic pits. The experiments and simulations presented below describe the possible detection of membrane curvature through fluorescence recovery after photobleaching (FRAP), fluorescence correlation spectroscopy (FCS), single particle tracking (SPT), and polarized localization microscopy (PLM). FRAP and FCS depend on diffraction-limited illumination and detection. In particular, a simulation of FRAP shows no effects on lipids diffusion due to a 50 nm diameter membrane bud at any stage in the budding process. Simulated FCS demonstrated small effects due to a 50 nm radius membrane bud that was amplified with curvature-dependent lipid mobility changes. However, PLM and SPT achieve sub-diffraction-limited resolution of membrane budding and lipid mobility through the identification of the single-lipid positions with ≤15 nm spatial and ≤20 ms temporal resolution. By mapping the single-lipid step lengths to locations on the membrane, the effects of curvature on lipid behavior have been resolved.

**Keywords:** fluorescence recovery after photobleaching; fluorescence correlation spectroscopy; single-particle tracking; supported lipid bilayers; membrane curvature engineering; diffusion; molecular shape;

## 1. Introduction

Biological membranes are commonly modeled as two-dimensional fluid bilayers composed primarily of phospholipids. While this approximation reproduces many of the key membrane features and complexities, lateral heterogeneity in composition and topography are increasingly recognized as fundamental elements critical for complex biological functions. The shape of biological membranes is precisely controlled for diverse, essential, cellular processes such as regulating organelle morphology, exocytosis/endocytosis, pathogen vulnerability/protection, and effective therapeutic targeting [1–3]. Accordingly, the dysregulation of membrane curvature is broadly implicated in cardiovascular disease, viral infections, cancer, Alzheimer's disease, Huntington disease, diabetes, and other diseases [4–6]. Each of these processes requires the fusion or fission of <50 nm radius vesicles with otherwise near planar membranes via precise regulation of the local curvature-generating forces [7]. The underlying non-specific, lipid-based influences remain relatively unknown, despite extensive evidence of their importance [8–10]. Nanoscale curvature is among the most important and least understood properties of biological membranes.

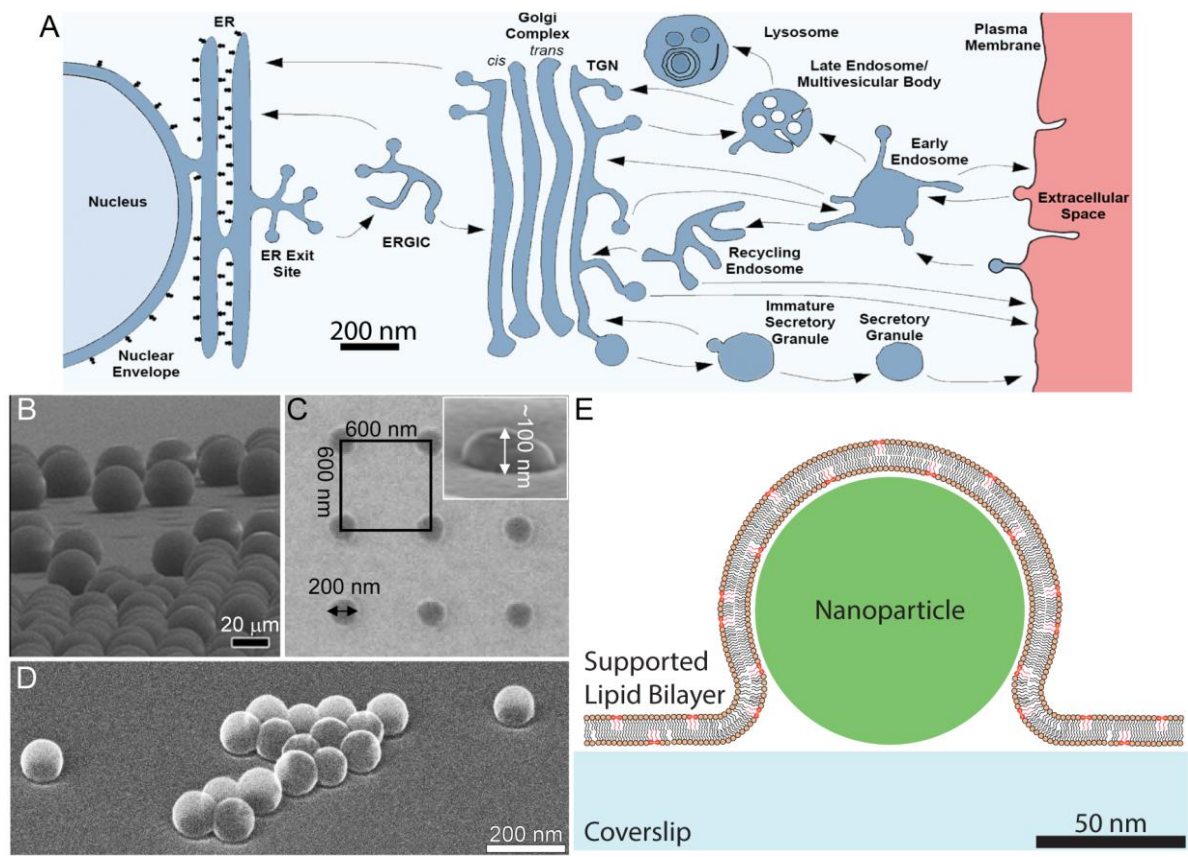

**Figure 1.** (A) The transport of material through the cell depends on precisely controlled membrane shapes. The nanoengineering of membrane shape analogous to membrane budding has employed supported lipid bilayer over (B) micron-scale particles, (C) electron-beam lithography, and (D) 50 nm diameter nanoparticles. (B-D) The substrate topographies without a membrane were imaged with SEM. (E) The final engineered membrane curvature includes a top that is determined by the nanoparticle size, and smooth connection to the surrounding planar supported lipid bilayer. (A), (B), and (C) have been adapted with permission from [3], [11] and [12], respectively.

As a first stage of understanding the ramifications of curvature, only on lipid sorting and dynamics, membrane curvature has been nanoengineered in model systems to isolate the interplay between membrane curvature and lipid biophysics. The two best-studied geometrical and structural shapes are those of tubular and budding membranes. Membrane tubules are naturally seen in organelle shape and diverse plasma membrane externalization and internalization behaviors. Membrane tubules have been engineered through microbead pulling [8], protein crowding [13], or molecular motor function from giant membrane vesicles [14]. However, the dynamic nature of the tubules has limited the nanoscale observation of lipids within them. Membrane buds are most commonly associated with the formation of endocytic pits preceding vesiculation. Further, buds are also relevant to the post-fusion state of exocytosis, in particular, kiss-and-run exocytosis [15]. Membrane bud engineering typically occurs on a supporting substrate, which enables high stability for super-resolution imaging of a supported lipid bilayer (SLB) draped upon a curved surface (Figure 1). For example, membrane buds were engineered with polystyrene microparticles to reveal the lipid phase and bud-neck association of cholera toxin [11]. Nanoscale membrane buds have been formed over polydimethylsiloxane (PDMS) curvature shaped via electron-beam lithography [12]. With these buds, the interplay of curvature and lipid phase has been clearly demonstrated [16,17]. Recent studies have combined the ease of fabrication with polystyrene beads and nanoscale dimensions of electron beam lithography, by draping SLBs over <50 nm radius polystyrene nanoparticles to reveal single-lipid dynamics [18–20] and curvature-based protein sorting [21].

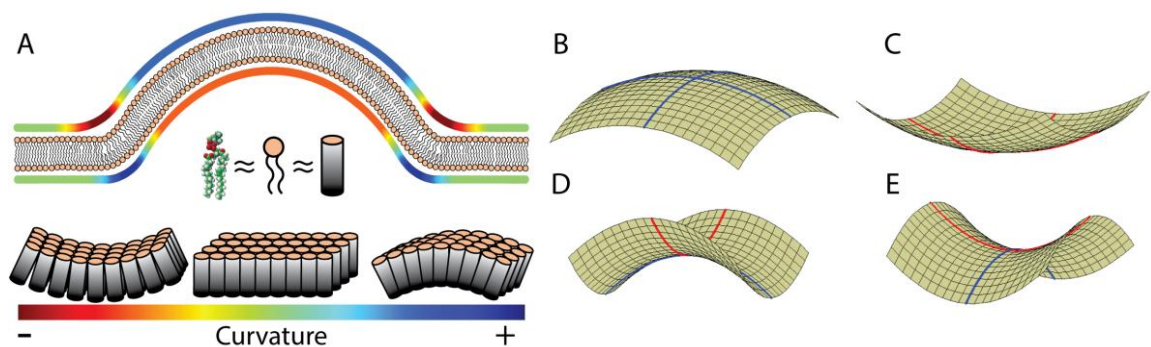

**Figure 2.** The coupling between curvature and lipid mobility is a critical component of membrane function. (A) Membrane bending requires the opposing leaflets within the bilayer to experience curvature of opposite sign along the principle component planes, such as the slice through the *xz*-plane as shown here. However, both leaflets to have a (B, C) positive Gaussian curvature on the bud top and a (D, E) negative Gaussian curvature at the bud neck as the 2D membrane surface and both principle curvatures are considered for any point on the membrane.

Within the membrane bud, there is both positive and negative curvature that complicates understanding the correlation between shape and lipid dynamics (Figure 2). Positive curvature is defined as when the lipid head groups are convex, and negative curvature is defined as when the lipid head groups are concave. A positive curvature results in an increased packing of the lipid tails while a negative curvature results in an increased packing of the lipid head groups. The molecular structure of the constituent lipids affects the intrinsic membrane shape and can lead to a sorting of lipids across the topography. Further, the Gaussian membrane curvature is defined as the product of the two principal curvatures at any point on the surface. For example, both the top and bottom leaflets of the bilayer at the bud top have positive Gaussian curvature, while the bud neck has a negative Gaussian curvature. Resolving the effects of curvature on lipid dynamics will require separately resolving the bilayer leaflets, as well as distinguishing the contributions of lipids on the neck-vs-top of the bud.

The submicron length scale and submillisecond time scale of lipid dynamics on curved membranes have made it experimentally challenging to determine their underlying biophysics on membrane curvature. Optical techniques enable the observation of living samples (*i.e.*, aqueous, 0 to 45 °C, and non-ionizing radiation). Interference contrast microscopy and polarized total internal reflection fluorescence microscopy (TIRFM) have enabled the detection of membrane curvature in diverse samples [22–31]. However, these optical techniques are traditionally limited by diffraction to a spatial resolution of >200 nm. Nanoscopic optical methods such as direct stochastic optical reconstruction microscopy [(d)STORM], photoactivated localization microscopy [(f)PALM], and stimulated depletion emission microscopy (STED) [32–36] have improved the resolution of optical microscopy to >10 nm. Super-resolution fluorescence localization microscopies have been adapted to yield fluorophore height or orientation by manipulating the fluorescence emission path, but at the cost of reducing the lateral localization precision [37–44].

We have developed polarized localization microscopy (PLM) to reveal dynamic, nanoscale, membrane curvature. Polarized localization microscopy combines single-molecule localization microscopy (SMLM) and polarized TIRFM [18]. Polarized TIRFM reveals membrane orientation by selectively coupling linearly polarized fluorescence excitation with lipidated indocarbocyanine dyes (*i.e.*, DiI, DiO, DiD) that maintain their fluorescence dipole moment in the plane of the membrane (Figure 3) [45]. P-polarized excitation preferentially excites the dyes in the membrane that have a transition dipole perpendicular to the coverslip; s-polarized excitation preferentially excites the dyes in the membrane that have a transition dipole parallel to the coverslip. There are two primary limitations of polarized localization microscopy: the fluorophore localizations accuracy and the rate of localizations, both of which contribute to limiting the resolution of the local membrane angle.

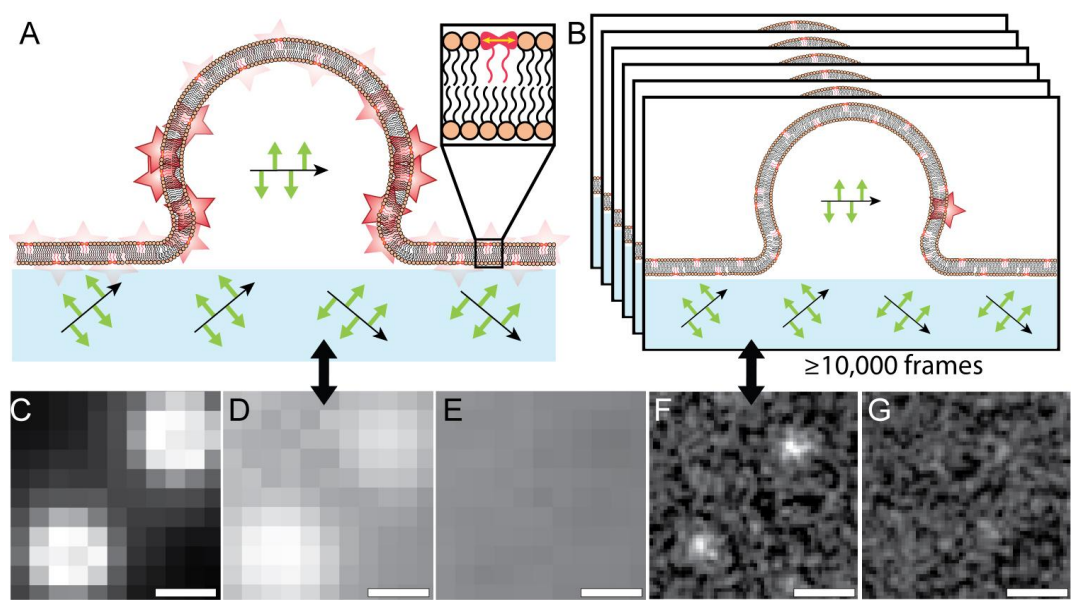

**Figure 3.** Polarized localization microscopy utilizes linearly polarized TIRF excitation and fluorophores that maintain their orientation within the membrane to yield an order-of-magnitude improvements in the sensitivity and resolution of membrane curvature. (A) P-polarized TIRF excitation preferentially excites DiI molecules within the membrane that is perpendicular to the coverslip. (B) The blinking of DiI with >50 mW of excitation light can be imaged with many sequential frames for individual fluorophore fitting and image reconstruction. (C) 51 nm radius fluorescence polystyrene nanoparticles were used to engineer membrane curvature, as imaged here with $\lambda_{ex}$ = 405 nm. (D-G) Polarized TIRFM with $\lambda_{ex}$ = 561 nm images the DiI within the curved bilayer. (D, F) P-polarized excitation emphasizes the vertical membrane. (E, G) S-polarized excitation emphasizes the horizontal membrane. The (D, E) diffraction-limited polarized TIRFM images provide an order-of-magnitude worse spatial resolution and signal-to-noise than (F, G) the PLM images. (C-G) Scale bar represents 250 nm.

As a pointillist imaging method, PLM provides raw data that can be interpreted for high-throughput single particle tracking of lipid diffusion dependent on membrane curvature. Tracking individual fluorophores (*e.g.*, DiI) that stay *on* for multiple sequential frames enables the observation molecules diffusion rates *versus* membrane topology. For example, DiI molecules diffuse on curved membranes at 25% of the speed at which they diffuse on flat membranes (Figure 3). This slowed diffusion was assessed in consideration of the geometrical contributions to the apparent diffusion through the *xy*-plane, which is only feasible with super-resolution analysis of the membrane topography. Analysis of single-molecule diffusion rates relative to membrane topology reveals information regarding the local environment (*i.e.*, lipid phase or molecular crowding) associated with membrane bending.

In this manuscript, we demonstrate the capabilities of various fluorescence techniques to reveal lipid dynamics relative to membrane curvature. We focus on the three most common methods of measuring lipid mobility: fluorescence recovery after photobleaching (FRAP), fluorescence correlation spectroscopy (FCS), and single-particle tracking (SPT). Through Monte Carlo simulations of Brownian diffusing lipids over membrane buds of varying heights, we demonstrate the ability of each of these techniques in revealing the presence of the membrane bud, the lipid dynamics on the bud, and the possible effects that an altered lipid mobility would have on the results. Our simulations demonstrate how FRAP was not sufficiently sensitive to reveal that a bud was present under any of our simulation conditions. FCS was able to reveal the bud's presence if the bud induces a slowed lipid diffusion, but FCS is typically limited to diffraction-limited length scales. SPT, however, was able to measure the effects of membrane topography changes with or without curvature-induced changes to lipid mobility. By mapping the single-lipid steps over space, buds of varying heights and membranes of laterally varying viscosity could be distinguished. Through carefully chosen methods, SPT data can reveal spatial information across the sample with <20 nm resolution. SPT can also indicate the presence of distinct populations of diffusing species. The effects of membrane curvature engineering, such as membrane-substrate adhesion, are further discussed in efforts to guide experimental design and advancing biophysical understandings.

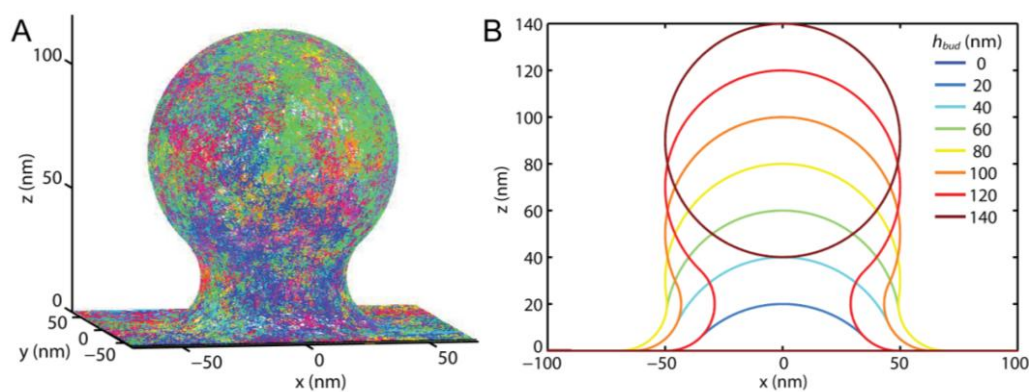

**Figure 4.** Simulations were performed by mimicking distinct stages of the fission and fusion of a 50 nm radius vesicle with a planar membrane. The bud top had a radius of curvature equal to 50 nm and was smoothly connected to the surrounding planar membrane with no less than 20 nm radius of curvature. The stage of the bud formation was characterized by the bud height ($h_{bud}$). (A) A static membrane topography with $h_{bud}$ = 120 nm shows the narrow bud neck. Single-lipid trajectories over this topography are shown in assorted colors to demonstrate the dense sampling of the membrane by these simulations. (B) The cross-section of the 3D membrane topography through the $xz$-plane reveals the increase in membrane area with increasing $h_{bud}$ and the separate vesicle from the planar membrane when $h_{bud} \geq 140$ nm.

## 2. Materials and Methods

The diffusion of lipids through membrane buds was simulated and analyzed to mimic the expected experimental results that would be obtained by a variety of fluorescence-based methods. With custom MATLAB (MathWorks, Inc.) programming, membrane buds were modeled with a radius of curvature equals to 50 nm and varying heights above a surrounding planar membrane ($h_{bud}$). The bud membrane was smoothly connected to the surrounding planar membrane with a radius of curvature equals to 20 nm along the principal plane radial from the bud center (Figure 4), as done previously [18,21]. $h_{bud}$ = 0 represents the case of a planar membrane with no bud protrusion. When $h_{bud}$ = 140 nm, the bud had detached from planar membrane such that there was no diffusion between the vesicle and the planar membrane.

Trajectories of the individual lipids were simulated upon the budding topography via a Monte-Carlo method for a discrete set of randomly distributed points. The discrete points were created at a constant density of 4 points/nm$^2$ across the bud top, the bud-to-planar membrane neck, and the surrounding planar membrane. At each time step, the lipid moved to one of the 50 ± 7 random points within 2 nm. This resulted in an average single step distance of 1.3 nm. To mimic the diffusion coefficient ($D$) of 1 µm$^2$/s over many steps, each time step would correspond to 1.1 µs. The trajectory of each lipid started at 1 µm away from the bud center, then diffused randomly upon the simulated membrane until it was >1µm away from the bud center. More than 10$^5$ different trajectories were simulated for each condition, and 1300 ± 100 of those trajectories made it onto the membrane bud for each $h_{bud}$. The methods of analyzing these trajectories were designed to mimic experimental fluorescence techniques.

Previous analysis of lipid mobility over nanoscale membrane buds found that $D$ upon the planar membrane ($D_{Plane}$) was 7.6x faster than that on membrane bud ($D_{Bud}$) [18]. To mimic the curvature-induced slowing of the lipid diffusion, the effective time per simulation step was changed to be 8.4 µs for each 1.3 nm step whenever the simulated lipid was above the surrounding planar membrane to mimic $D_{Bud}$ = 0.13 μm²/s while $D_{Plane}$ = 1 μm/s². Simulations were performed with and without this curvature-induced slowing of the lipid diffusion.

All of the analyses done in this manuscript limits the observation to the *z*-projection of the fluorescence signal into the imaging *xy*-plane. The fluorescence emission was assumed to have no *z*-dependence or polarization dependence, as would be expected for epifluorescence illumination and randomly tumbling fluorophores. This assumption holds well when applied to the nanoscale structures that vary by less than 150 nm in the *z*-direction, as done here. Super-resolution methods such as PLM incorporate a polarization and *z*-dependence in the excitation, and may yield a varying localization probability *versus* membrane height or orientation. However, these effects in SPT from PLM are negligible compared to experimental uncertainty with s- and p-polarized illumination providing indistinguishable SPT results [18].

*2.1 Mimicking Fluorescence Recovery After Photobleaching*

Fluorescence recovery after photobleaching (FRAP) measures the recovery of a fluorescence signal from a region of the sample after all the fluorophores within that region are bleached when time (*t*) equals to zero. Here, the bleached region was 2 μm in diameter. The time constant of the signal recovery ($\tau_{FRAP}$) of 0.3 sec was expected, and a fluorescence recovery signal that lasted for 0.9 sec was needed to achieve adequate recovery to steady-state intensity. A spatially uniform fluorescence illumination was simulated such that the number of lipids present on the membrane was proportional to the fluorescence signal. For simulating the intensity (*I*) versus *t*, a new single-lipid trajectory was started at the perimeter of the region of interest every 15 μs. All lipid trajectories contributed to the sum signal until the trajectory moved more than 1 μm away from the membrane bud. The sum of the simultaneously present lipid trajectories was then interpreted as the *I* versus *t* signal for the recovery of the fluorophores into the region of interest after photobleaching and fitted to

$$I_{FRAP\ Fit}(t) = A(1 - e^{t/\tau_{FRAP}}). \tag{1}$$

The fitting variable *A* represents the steady-state magnitude of *I* and is proportional to the steady-state fluorophore density, illumination intensity, and fluorescence emission collection efficiency. Commonly the FRAP fitting is further analyzed to consider an immobile fraction, but this was not necessary for these simulations.

*2.2 Mimicking Fluorescence Correlation Spectroscopy*

Fluorescence correlation spectroscopy (FCS) examines the fluctuations in the steady state *I* versus *t* signal through calculating the autocorrelation (*G*) as a function of lag time (τ) and finding the characteristic fluctuation time ($\tau_{FCS}$). In these simulations, *I(t)* for FCS was calculated from the single-molecule trajectories (*x(t)* and *y(t)*) through a Gaussian illumination profile according to

$$I_{FCS}(t) = \exp(-(x(t)^2 + y(t)^2)/w^2), \tag{2}$$

with an illumination width (*w*) was set equal to 200 nm, as would be expected for typical confocal FCS. *G* was calculated from *I(t)* according to

$$G(\tau) = <\delta I(t)\delta I(t-\tau)>/<I(t)>^2. \tag{3}$$

The angle brackets (<>) represent the average over *t* and $\delta I(t) = I(t)-<I(t)>$. The correlation time ($\tau_{FCS}$) in *I(t)* was found by fitting *G(τ)* according to

$$G_{Fit}(\tau) = G_0(1 - (\tau/\tau_{FCS})^2)^{-1}, \tag{4}$$

as is expected for 2D Brownian diffusion. The fitting variable $G_0$ is inversely proportional to the number of diffusers simultaneously observed and the constant background contribution to *I*. With a membrane bud present, the autocorrelation is not expected to fit perfectly to Eq. 3; however, the inherent averaging incorporated into an autocorrelation analysis makes finding minor populations difficult, and complex fitting functions are typically unwarranted [46]. For this analysis, Eq. 2 was assumed to be both the spatial illumination and detection sensitivity, as would be expected for diffraction-limited, single-spot analysis.

*2.3 Mimicking Single Particle Tracking*

Single-particle tracking (SPT) includes identifying the center of each single-fluorophore image via computational analysis of a movie of sparse, dynamic fluorophores. From the motion of the single-molecules between sequential frames, a single-molecule trajectory was observed. Typically, the trajectories analyzed by calculating the mean squared displacement (*MSD*) versus lag time (*Δt*) such that 2D Brownian diffusion results in a linear relationship of

$$MSD(\Delta t) = 4D\Delta t + 2\sigma_r^2 - 8DRt_{frame}. \tag{5}$$

The 2D localization uncertainty ($\sigma_r^2 = \sigma_x^2 + \sigma_y^2$) and camera blur contribute to these last two terms of Eq. 5, respectively. Camera blur depends on the time between adjacent frames ($t_{frame}$) and the motion blur coefficient (*R*). *R* depends on the camera exposure duration ($t_{exp}$) such that $R = t_{exp}/(6t_{frame}) \leq 1/6$ for continuous exposures within each frame [47]. Typically, only the *Δt* = 2$t_{frame}$ through *Δt* = 4$t_{frame}$ are used, but diverse fitting conditions have been previously used to extract particular sample details, including confinement effects at longer length and time scales [48]. However, this MSD *versus Δt* analysis spatially averages each trajectory. For trajectories that include dozens of sequential localizations, this analysis may provide an average over many square microns of the sample, well beyond the extent of a membrane bud.

Alternatively, SPT analysis has been performed with a single-step length analysis such that regions of shorter steps can be associated with localized regions of membrane bending and/or a locally varying membrane viscosity [18,21]. The single-steps lengths (*s*) for any region of the sample may be fit to a 2D Maxwell-Boltzmann or Rayleigh Distribution to determine the local *D* where the probability distribution of step length for a single Brownian diffuser in a uniform membrane (*P*) is

$$P(s) = \frac{s}{2D\Delta t} e^{-\frac{s^2}{4D\Delta t}}. \tag{6}$$

As shown below, the observed single-molecule steps were grouped and fitted according to their distance from the bud center (*r*) such that *D versus r* could be measured.

However, the single-step analysis with the experimental limitations that $\sigma_r > 0$ and $t_{exp} > 0$ yield a systematic difference between the *D* found from fitting Eq. 6 ($D_{Fit}$) and the *D* that would be found from a perfect experimental system ($D_{Real}$) if $\sigma_r$ and $t_{exp}$ approached zero, according to

$$D_{Real} = (D_{Fit} - \frac{\sigma_r^2}{2\Delta t})/(1 - \frac{t_{exp}}{3\Delta t}). \tag{7}$$

A quicker frame rate (*i.e.*, smaller $t_{frame}$) typically yields a worse signal-to-noise ratio in the raw data (*i.e.*, higher $\sigma_r$) that both contribute to effects of the differences between $D_{Fit}$ and $D_{Real}$ in this single-step analysis. However, in the SMLM systems, a frame rate up to 500 Hz typically allows for greater spatial resolution of the sample viscosity, since the tracked lipids diffuse less distance between localizations, as discussed below. Similarly, a blur of the single-molecule image occurs if the single frame acquisition time is comparable to the time between frames (*i.e.*, $t_{exp} \approx t_{frame}$). However, the effect of camera blur diminishes as nonadjacent frames are compared and a larger *Δt* is used, but this is performed at the sacrifice of total distance diffused by the molecule during the single-steps analysis.

When the membrane is not parallel to the coverslip, then the *z*-component to the lipid diffusion within the membrane results in an apparent slowing of the lipid through the *xy*-plane. It was not possible to extract the in-plane diffusion rate from the observed diffusion through the *xy*-plane ($D_{xy}$) when both the membrane topography and the influence of curvature on membrane viscosity are unknown. In the below analysis, $D_{real}$ was calculated under the approximation that the membrane was parallel to the coverslip (Eq. 7), and this value was reported as $D_{xy}$ to be explicit in representing that the membrane topography was contributing to the measurements.

*2.4 Sample preparation*

For experimental observation of the nanoscale membrane topography, glass bottom dishes (MatTek Corp.) were cleaned by immersion in 7x detergent overnight, rinsed with 18.2 MΩ-cm water (Milli-Q, EMD Millipore Corp.), bath sonicated for 30 min, dried with nitrogen gas, and cleaned by air plasma (Harrick Plasma). Nanoparticle sedimentation occurred for 10 min to achieve an average density of 0.02 NPs/µm$^2$, as described previously [18]. Dishes were placed on a 55 °C hot plate for 5 min to ensure their stability on the coverslip. Supported lipid bilayers over the nanoparticles were formed by the fusion of giant unilamellar vesicles of primarily 1-palmitoyl-2-oleoyl-sn-glycero- 3-phosphocholine (POPC, Avanti Polar Lipids, Inc.) labeled with 0.3 mol% 1,1'-didodecyl-3,3,3', 3'-tetramethylindocarbocyanine perchlorate (DiI, Life Technologies). Giant unilamellar vesicles were prepared by electro-formation, as described previously [49].

PLM was performed on samples present in an oxygen-scavenging buffer (150 mM NaCl, 50 mM TRIS, 0.5 mg/mL glucose oxidase, 20 mg/mL glucose, and 40 µg/mL catalase at pH 8). Buffer proteins were purchased from Sigma-Aldrich and salts were purchased from Fisher Scientific. These conditions maintain a low free oxygen concentration in the buffer to minimize non-reversible fluorophore bleaching and encourage transient fluorophore blinking, as is necessary for SMLM.

*2.5 Optical and electron imaging*

PLM and SPT were performed with an inverted IX83 microscope with zero-drift correction and a 100x, 1.49NA objective (Olympus Corp.) on a vibration-isolated optical table, as described previously [18]. The excitation polarization was rotated with computer-controlled liquid crystal waveplate (LCC1111-A, Thorlabs Inc.). Image acquisition was performed with an iXon-897 Ultra EMCCD camera (Andor Technology) proceeded by emission filters (BrightLine single-band bandpass filters, Semrock, Inc.), a 4-band notch filter (ZET405/488/561/640m, Chroma Corp.), and a 2.5x magnification lens (Olympus Corp). Exposure of the sample to >80 mW of excitation light with $\lambda_{ex}$ = 561 nm for 3 sec resulted in converting most of the DiI from the fluorescent state (*'on'*) to the transient, non-fluorescent, dark state (*'off'*) to provide steady state fluorophore blinking. The *'on'* fluorophores were imaged at a density of less than one fluorophore per 1 µm$^2$. Sequential movies were acquired with alternating p-polarized TIRF (pTIRF) excitation at $\lambda_{ex}$ = 561 nm for pPLM and s-polarized TIRF (sTIRF) excitation at $\lambda_{ex}$ = 561 nm for sPLM. 10,000 to 30,000 frames were acquired for each polarization at a frame rate of 50-500 Hz.

The Fiji plug-in ThunderSTORM was used to calculate the locations of the fluorophores from each acquired frame [50,51]. We kept only the localizations with ≥10 detected photons and $\sigma_r \leq 45$ nm for further analysis. The sequential locations of the same fluorophore were linked by u-track [52]. MSD of each trajectory was calculated as a function of maximum searching radius. When the MSD remained the unchanged for a range of maximum searching radii, there were minimal missed links of single fluorophores and minimal false linking between different fluorophores.

A field emission scanning electron microscopy (SEM, JSM-7600F, Jeol USA, Inc.) was performed in the Wayne State University Electron Microscopy Laboratory. As done for the optical measurements, 51 nm radius polystyrene nanoparticles (FluoSpheres, Thermo Fisher Scientific Inc.) were sedimented onto a glass coverslip and melted on a hot plate at 55°C for 5 min to ensure stability and encourage adhesion to the substrate. The nanoparticles were carbon coated and imaged at an angle of 55° with a secondary electron detector to reveal the effects of the hotplate on the nanoparticle shape.

## 3. Results

Substrate nanoengineering through the application of 51 nm radius polystyrene nanoparticles, provides an experimental method for constructing membrane topography of physiologically relevant dimensions to mimic clathrin- or caveolin-mediated endocytosis. The scanning electron micrograph of the nanoparticles shows that even after 5 min of melting at 55 °C, the nanoparticles maintain their spherical structure (Figure 1C). Bursting giant unilamellar vesicles over the nanoparticles and the supporting coverslip provided a continuous connection between the engineered membrane bud and the surrounding planar supported lipid bilayer (Figure 1B). This membrane engineering approach created the positive and negative membrane curvature as expected during endocytosis and exocytosis (Figure 1A) and was readily apparent with PLM (Figure 3). With the ability to engineer the membrane topography, detect it with super-resolution optical methods, and employ various techniques to reveal lipid diffusion, simulations have been used to predict how FRAP, FCS, and SPT could be optimized to provide biophysical understandings.

*3.1. PLM detection of membrane bending*

The use of PLM to detect membrane bending over engineered substrates enables order-of-magnitude improvements in the spatial resolution and detection sensitivity of membrane curvature as compared to other optical methods. Membrane curvature was engineered with 50 nm diameter polystyrene nanoparticles and supported lipid bilayers composed of 99.4 mol% POPC were created over the nanoparticles (Figure 3). The nanoparticles were detected by their fluorescence with an excitation wavelength ($\lambda_{ex}$) of 405 nm and emission at wavelengths between 420 and 480 nm (Figure 3A). DiI within the membrane was detected with $\lambda_{ex}$ = 561 nm and emission at wavelengths between 570 and 620 nm. The linearly polarized total internal reflection fluorescence excitation enables the vertical and horizontal electric fields to preferentially excite DiI in the membrane that was perpendicular or horizontal to the coverslip, respectively. Polarized TIRFM was typically performed with diffraction-limited resolution. However, by using >50 mW of excitation light and an oxygen scavenging buffer, the DiI was made to blink for single-molecule observation. The single-molecule localizations were analyzed for image reconstruction (Figure 3D, E) or SPT to reveal the effects of curvature due to membrane bending, as described below.

*3.2. Membrane bending and FRAP*

Experimental FRAP conditions such as the laser-based illumination with intense fluorophore bleaching and small observation areas were mimicked in the simulations presented here. A complete bleaching of the fluorophores in a 2 μm diameter spot and a uniform illumination during recovery were simulated. Accordingly, the location of the bud within the observation area did not affect the results. Further, FRAP results did not change significantly upon bud formation even if the membrane curvature induces slowed lipid diffusion (Figure 5).

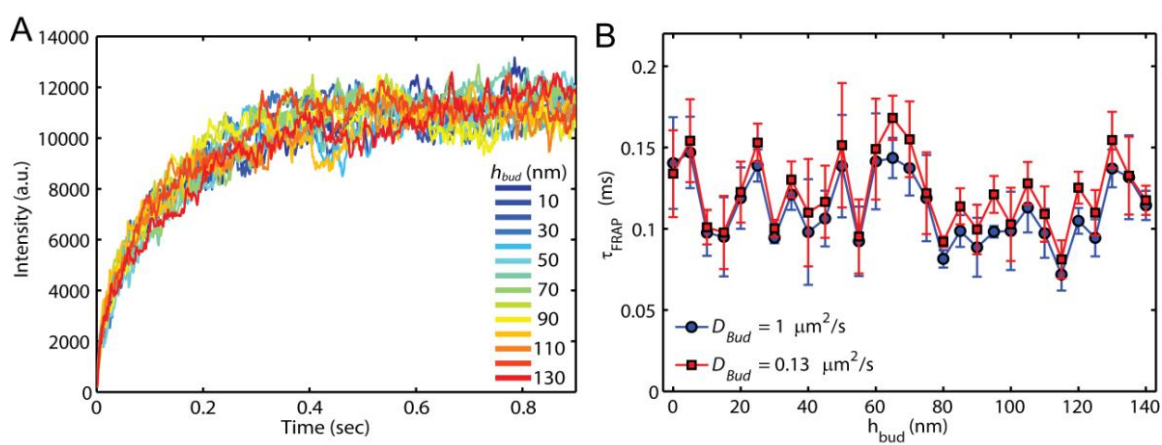

**Figure 5.** Increasing the bud height does not result in significant changes to the FRAP recovery time, including when the lipid diffusion was slowed to 0.13 μm²/s on the curved membrane. (A) $I(t)$ traces while $D_{Plane}$ = 1 μm²/s and $D_{Bud}$ = 0.13 μm²/s shows the recovery of $I(t)$ after bleaching at $t$ = 0. There was no apparent trend in the recovery rate changing with bud height in the 2 μm diameter observation spot. (B) The recovery rate was quantified by fitting Eq. 1 to find $\tau_{FRAP}$ from $I(t)$ of each condition. Error bars demonstrate the standard error of the mean between separately processed thirds of the all simulated trajectories.

*3.3. Membrane bending and FCS*

The use of correlation functions by FCS enables the detection of minor changes to the lipid mobility. FCS can be performed with hardware or software-based correlators, and $G(\tau)$ requires minimal processing prior for interpretation; however, FCS requires laser illumination and a sub-millisecond temporal resolution in the $I(t)$ data. The presence of a single nanoscale membrane bud can be detected with FCS with or without curvature-induced lipid slowing (Figure 6). Even with diffraction-limited illumination (Eq. 2) and imprecise centering of the observation spot over the membrane bud, FCS was able to detect the effects of the bud on the lipid dwell time. For example, the effects of the membrane bud are on average 20% less when the center of the membrane bud was 100 nm offset away from the center of the diffraction-limited FCS illumination spot (Figure 6B).

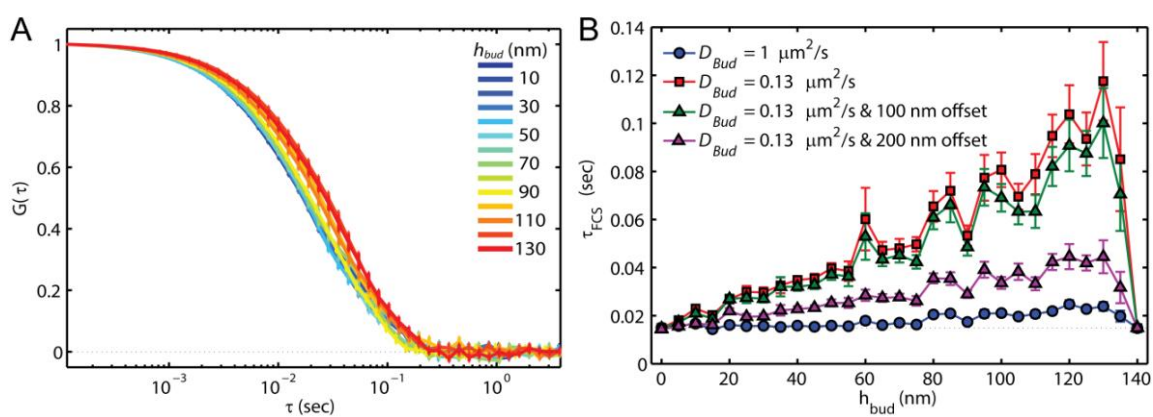

**Figure 6**. A membrane bud may be detected by FCS if $h_{bud}$ > 80 nm or the membrane curvature slows the local lipid diffusion. While a lipid on the planar membrane experiences $D_{Plane}$ = 1 μm²/s, varying the diffusion rate of the lipid on the curved membrane and the location of the bud within the excitation spot affects the observed FCS results. (A) A shifting of $G(\tau)$ to longer lag times was apparent when $D_{Bud}$ = 0.13 μm²/s and the bud was centered on the 200 nm wide Gaussian illumination. (B) If $D_{Bud}$ = 1 μm²/s and $h_{bud}$ < 80 nm, no effect of the bud was observed; the extra membrane area and varying membrane orientation are not sufficient to affect the FCS results. However, if $D_{Bud}$ = 0.13 μm²/s, the presence of the bud becomes clear for $h_{bud}$ > 15 nm. The bud can be offset by 200 nm and still yield a clear change in the dwell time through the observation spot. If the observation spot was off-centered from the bud by 100 nm or 200 nm, the effects of the bud are decreased by an average of 20% or 70%, respectively.

*3.4. Membrane Bending and SPT*

The process of performing SPT of individual fluorophores requires similar experimental system as that needed for SMLM. Namely, >50 mW illumination and a single-fluorophore sensitive camera are necessary. Further, SPT requires similar peak-finding data analysis and trajectory linking computational analysis for interpretation. The balance of the total fluorophore density, the fluorophore '*on*' fraction, the collection rate of photons per fluorophore per second, the diffusion rate of the fluorophore, and the camera frame rate, all must be carefully balanced to ensure detection of the isolated fluorophores and proper linking of the single-fluorophore trajectories between adjacent acquisition frames. However, significantly more details of the fluorophore behavior across the sample can be extracted from SPT results than FCS or FRAP.

Mapping the locations of single molecule steps enables mapping *D* across the sample (Figure 7). When a membrane with consistent viscosity and consistent lipid mobility was simulated, the variations in the measured $D_{xy}$ across the sample was due to the membrane topology change, and the perceived shorter lipid step lengths on tilted membranes as the lipid trajectories are projected on the imaging *xy*-plane (Figure 7A). The effects of the membrane bud presence are enhanced when the curved membrane reduces the lipid mobility (Figure 7B). When the membrane bending occurs in a rotationally symmetrical way, as would be expected for an endocytic pit and the engineered curvature shown here, a radial averaging provides greater clarity in the effects of membrane budding on the SPT results (Figure 8).

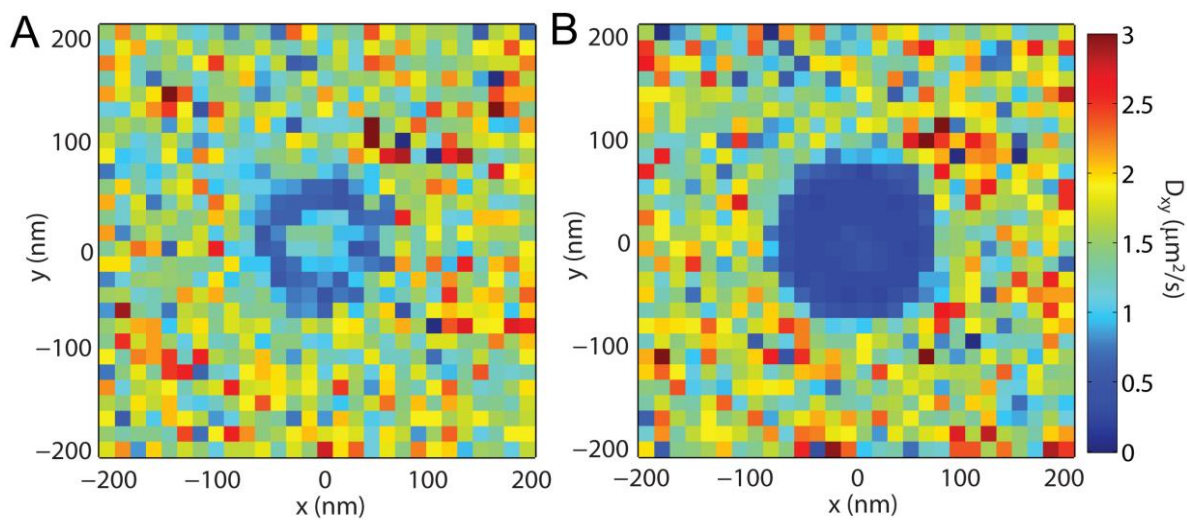

**Figure 7.** $D_{xy}$ mapped over the sample through SPT. The single steps from all trajectories are binned according to the average x position of the two localizations, and the average y position of the two localizations. This 2D binning of the *x* and *y* positions allows for the analysis of all the step lengths in a given region on the sample, as represented by a single pixel in these images. The azimuthal symmetry of the modeled membrane topology enables analysis of the observed diffusion rates versus distance from the bud center (Figure 8). Here, $h_{bud}$ = 100 nm, $\sigma_r$ = 15 nm, and $t_{frame}$ = 2 ms while (A) $D_{bud}$ = 1 µm²/s and (B) $D_{bud}$ = 0.13 µm²/s. The bud induced slowing in (A) was due to the membrane topography causing the lipid to move slower through the *xy*-plane with constant in-membrane diffusion.

## 4. Discussion

Resolving the nanoscale biophysical effects of membrane curvature remains experimentally challenging. Optical techniques such as PLM enable the detection of curvature with higher sensitivity and resolution than comparable optical techniques while providing a biologically friendly imaging environment. However, decoupling the effects of varying membrane area, orientation, and curvature can be challenging when the primary data collected was the *z*-projection and sample topography changes on sub-diffraction-limited length scales. Compensating for the effects of membrane topography on data analysis and interpretation was essential to reveal the effects of curvature.

*4.1. Engineering curvature at physiological length scales*

The diverse vesicular trafficking pathways within eukaryotic cells necessitate precisely controlled membrane curvature with radii of curvature ≤50 nm (Figure 1A). Macropinocytosis can create vesicles of micron-scale for the internalization of large particles, such as food or synthetic microspheres. As the pinocytotic pit forms and the membrane wraps around the particle, membrane curvature was tightly bent at select locations in the phagosome, namely at the end of the pseudopod and at the intersections between the pseudopod and the plasma membrane. In the engineering of micron scale curvature, the ≥20 µm diameter microspheres displayed curvature-dependent membrane organization and dynamics at the tightly bent bud neck with a radius of curvature an order-of-magnitude smaller than the top of the bud [11].

The engineering of curvature at the scale of clathrin and caveolin-mediated endocytosis requires reducing the bud size by an order-of-magnitude. In these nanoscale receptor-mediated processes, the membrane composition and curvature are significantly changed throughout the endocytotic pit. Electro-beam lithography and nanoparticle-based lithography have been used to create the <100 nm diameter pits and reveal lipid phase separation [12,16,17], membrane viscosity changes [18], and molecular sorting effects [19,21]. Electron-beam lithography provides the benefits of greater flexibility in the bud dimensions and the distribution of sizes across a single sample. However, electron-beam lithography requires expensive instrumentation in a low-particulate cleanroom. Nanoparticle-based substrate engineering provides the benefits of straightforward wet-chemistry fabrication that can be done in most academic research labs. The nanoparticle-based fabrication results in variations across a sample, bud-to-bud distance, and nanoparticle aggregation. However, when the nanoscopic details of sorting around isolated buds are the desired, select sample locations can be found on the coverslip with the desired features. Typically, 10 ± 5 isolated nanoparticles with a coating membrane were found in 100 μm² for PLM and SPT measurements. Additionally, the use of PLM enables verification that the nanoparticle-based membrane curvature engineering was accomplished (Figure 3).

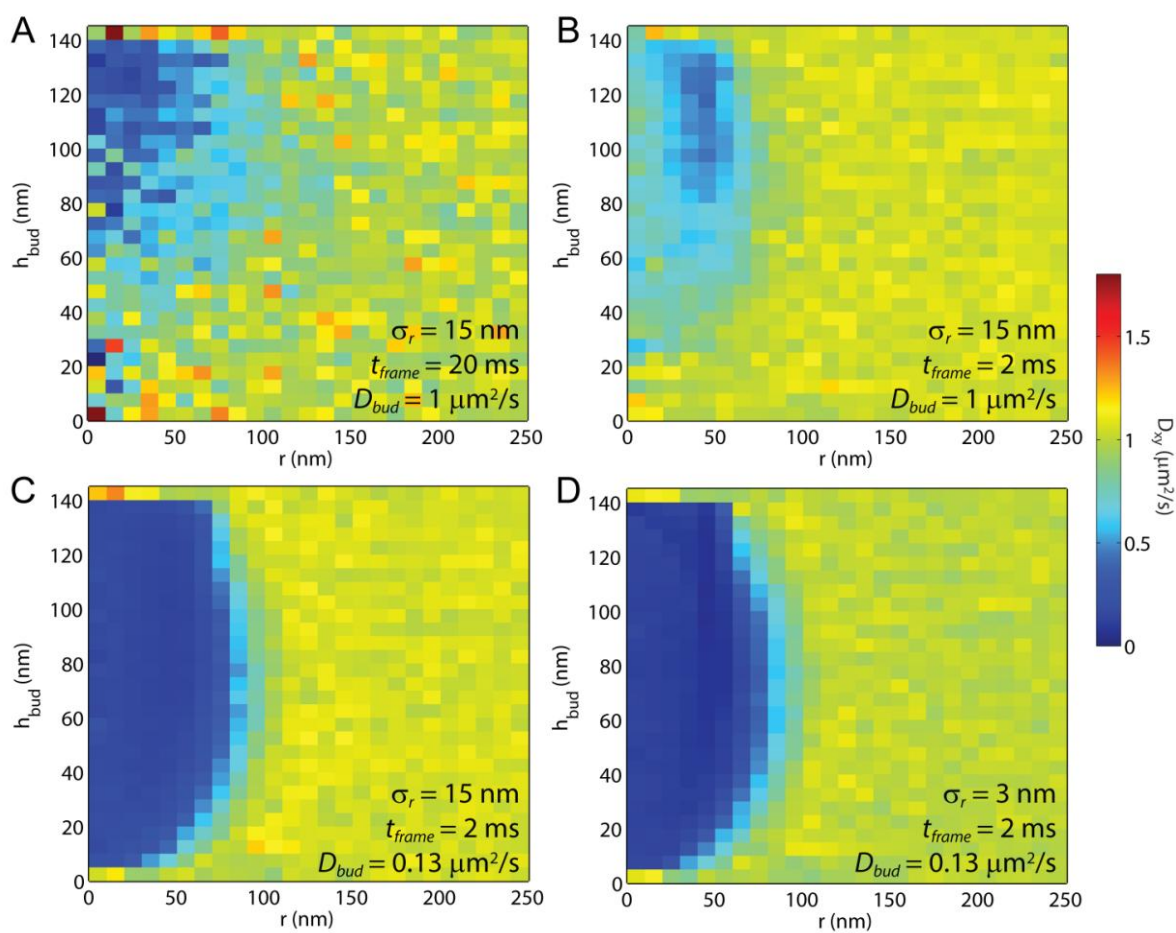

**Figure 8.** Azimuthal averaging of the spatial mapping of the $D_{xy}$ around a bud (*i.e.*, Figure 7) improves the data statistics and enables easier comparison of $D_{xy}$ versus distance from the bud center (*r*) and $h_{bud}$. Simulated SPT results for $D_{xy}$ are presented with varying $\sigma_r$, $t_{frame}$, and $D_{bud}$. (A) A long $t_{frame}$ results in significant data blurring and a slower data acquisition rate. (A, B) When $D_{bud} = D_{plane} = 1$ μm²/s, the single molecules maintain a uniform average local speed through the membrane, and the observed variations in $D_{xy}$ are due to the tilt of the membrane. (C, D) The membrane bud becomes sharply defined when $D_{bud} < D_{plane}$ and a fast image acquisition rate was used. (C, D) Improving the localization certainty has minimal effects compared to the acquisition frame rate.

*4.2. Comparative ease FRAP, FCS, and SPT*

FRAP, FCS, and SPT each provide benefits in regards to the specific membrane processes to which they are sensitive to, and the ease by which they are experimentally performed. FRAP is the easiest to these techniques in both the execution and analysis of the experiment. FRAP can be carried out on large observation regions with a conventional epifluorescence microscope through the opening and closing a field diaphragm in the conjugate image plane. Although this method of performing FRAP can provide a coarse analysis of membrane integrity and lipid mobility, it can be difficult to achieve a high intensity of the fluorescence emission and small enough observation areas to provide precise measurements. Even if performed with a relatively weak and slow bleaching procedure, FRAP is able to reveal the fraction of the diffusers that are immobile and the average diffusion coefficient of the mobile diffusers in a large observation area. This is especially valuable for demonstrating the continuity of a model membrane.

The correlation of intensity versus time for FCS only reports the diffusers that move through the diffraction-limited observation spot over the ~30 sec of data collection; FCS does not incorporate any information from immobile particles on a membrane. However, FCS is more sensitive to sub-populations of diffusers than FRAP and provides greater accuracy in the measured diffusion coefficients of the mobile diffusers. Commercial FCS setups can require expensive detectors, hardware correlators, and software, but are able to provide analysis of the results in real-time. Homebuilt FCS setups may use high-frame rate EMCCD or sCMOS cameras and custom software for correlation calculation and fitting. Since FCS is able to reveal late-stage bud formation and the effects of bending on membrane viscosity, it is feasible that future incorporations of FCS will be used to report the biophysical ramifications of membrane bending.

As shown above, SPT was able to provide the best spatial resolution of membrane bending and the effects of bending on membrane viscosity. However, SPT requires significantly more effort in data collection and analysis. SPT requires a high photon flux for precise single-fluorophore fitting ($\sigma_r$ <20 nm) with fast frame rates (≥50 Hz), which often requires oxygen-scavenging buffers to reduce fluorophore oxidization to provide more photons per fluorophore '*on*' state and greater conversation from the fluorophore '*off*' to '*on*' states. The raw SPT data typically comprises ≥5k individual camera images from which the single-molecule locations are calculated. The locations are then linked for trajectory analysis and MSD or single-step size analysis. A complicating factor of SPT is that the data quality and signal-to-noise ratio can vary between experiments such that user confirmation is needed for the analysis of each experiment. Despite these experimental challenges, the resolution benefits of SPT commonly justify its implementation.

*4.3. The diffraction limit and spatial resolution of mobility*

The method of draping a supported lipid bilayer over a nanoparticle has been used in prior experimental studies to reveal the influence of curvature on lipid dynamics and protein sorting *[18–21]*. FRAP has been successfully used to confirm the continuity of the membrane over the nanoparticles. However, FRAP was unable to reveal any difference in the recovery rate of the fluorescence due to the presence of curvature in experiments *[19]*, as expected by the simulations performed here (Figure 5).

Even with diffraction-limited illumination (Eq. 2) and imprecise centering of the observation spot over the membrane bud, FCS was able to detect the effects of the bud on lipid diffusion in these simulations (Figure 6). However, the ability for FCS to detect the membrane topology without curvature-induced lipid slowing was limited, and the ability for FCS to reveal the lipid mobility on different parts of the bud are prohibited by the size of the diffraction-limited illumination. By scanning over the sample, it would be feasible to provide sub-diffraction-limited resolution of the bud's location in the sample without a complimentary signal by finding where $\tau_{FCS}$ was most slowed. However, it would be difficult to confirm this slowing of the lipid mobility was due to a membrane bud and not another membrane defect, such as a rare membrane-substrate interaction without a colocalized fluorescent nanoparticle, accumulation of clathrin, or simultaneous atomic force microscopy, for example. Similar to single-particle localizations, the greatest effects on the FCS results would be observed when the bud was centered in the illumination profile. Potentially sub-diffraction-limited STED-FCS could yield a greater resolution of the lipid mobility on distinct parts of the bud as well as increase sensitivity to the membrane topography itself *[53]*; however, STED-FCS is technically challenging, expensive, and rare.

By its very nature, SPT provides sub-diffraction-limited spatial resolution. Like FRAP, SPT has been used to demonstrate the continuity of the SLB over the nanoparticle by visualizing single lipid trajectories that transition between the planar and curved regions of the membrane [18–20]. By binning the single-lipid step size versus distance from the center of the membrane bud and fitting the resulting histogram of step lengths to the Rayleigh distribution, the effects of lipid topography can be revealed directly even without curvature-affected lipid mobility (Figure 7, 8A). The effect of the membrane bending was observed $D_{xy}$ in these simulations without any curvature-induced changes to $D$. Similarly, observation of $D_{xy}$ across a sample could reveal a previously unknown membrane topography if there were curvature induced change to lipid mobility were known.

Individual fluorophore localizations are commonly made to ≤15 nm certainty. However, the spatial resolution of the mobility measurement is typically determined by the distance between sequential localizations rather than the precision of single localizations. When fluorescent trackers are used with more fluorescence emission than a single fluorophore, localization precision of <3 nm has been achieved [54]. However, the error in mapping mobility of a tracer to a location in the sample is typically limited by the distance the tracer has moved between sequential localizations. For example, a typical $t_{frame}$ on a sensitive EMCCD camera is 50 Hz. With a cropped-sensor mode, a reduced region of interest, and/or a modern sCMOS camera, single fluorophores can be detected at ≥500 Hz. With an expected mean distance between sequential localizations of a Brownian Diffusion equals to $(4D\Delta t)^{1/2}$, a frame rate of 50 Hz or 500 Hz can incorporate a spatial averaging of 280 nm and 90 nm, respectively, when $D = 1$ $\mu m^2/s$. This averaging can be of the same size as the membrane bud (Figure 8A). Further, the comparison of panels C and D of Figure 8 demonstrates how the improved localization precision resulted in no significant resolution or single-to-noise increase in the mapping of $D_{xy}$ versus $r$.

Recently, non-fluorescence, interferometric approaches have been demonstrated where gold nanoparticles have provided 1.7 nm localization precision with $t_{frame}$ = 1 ms for a single-step length of 60 nm between localizations of single lipids [55]. When pushed to extreme frame rates, the effects of localization precision can be the dominant source of uncertainty, including those with 17 nm localization precision with $t_{frame}$ = 0.03 ms that were used to detect hop diffusion in cell membranes [56]. However, these experiments with <1 ms frame rates depend on >20 nm diameter gold nanoparticle labels and are associated experimental uncertainties that are not present with single-fluorophore labels. The uncertainties associated specifically with gold nanoparticle labels include the reduced specificity of the number of lipids per nanoparticle, the effects of drag on the nanoparticle, the non-specific binding between the nanoparticle and the other membrane components, and the local heating that may be caused by the gold absorption of the illumination.

*4.4. PLM considerations*

PLM has enabled the detection of nanoscale membrane curvature with order-of-magnitude improvements in sensitivity and resolution over comparable techniques (Figure 3) [18]. The resolution of PLM is limited by both the rate of detected fluorophores and the localization precision of each fluorophore. The single-fluorophore detection requires a detection rate of ≤0.001 fluorophore/nm$^2$/s. With >10 fluorophores detected per bud required to reveal the presence of the bud, this yields a minimum bud detection time of 80 sec, which limits the temporal resolution of PLM.

While PLM and localization microscopy generally has $\sigma_r$ proportional to 1/sqrt(N) [57], PLM has the added uncertainty associated with the anisotropic emission of the rotationally confined fluorophores. As DiI molecule is able to spin in the membrane, but not tumble, and never point its emission dipole parallel to the membrane normal, it can yield a systematic inaccuracy in the localized position of the fluorophore. This is most enhanced when the membrane normal is tilted 45° to the coverslip normal and the membrane is >50nm out of focus. When a DiI molecule is 200 nm out of focus, the systematic shift can be up to 30 nm. For the membrane bud topography with the supported lipid bilayer in-focus, this results in all bud-associated DiI localizations being shifted towards the center of the bud and the diminishment of a ring of localizations, as would be expected from accurate localizations. This inaccuracy affects the reconstructed PLM images and the mapping of the fluorophore trajectory to a specific location on the bud. However, a theoretical analysis of PLM has incorporated these effects and demonstrated good agreement with the localizations from membrane buds created from nanoparticles of 24, 50, and 71 nm radius [18]. If precise analysis of the lipid diffusion was needed separately from the anisotropic effects of DiI, a second, chromatically distinct fluorescent label with could be used. For example, the excitation of dihexadecanoyl-phosphoethanolamine-rhodamine (DPPE-Rh, Avanti Polar Lipids, Inc.) displays no polarization or membrane-orientation dependence due to the flexible linkage between the rhodamine and the DPPE and may prove to be a better tracer for studying lipid diffusion. However, the polarization dependence and anisotropic emission effects of DiI seem to not affect the observed diffusion of the DiI through a membrane bud since p- and s-polarized illumination provides indistinguishable results [18].

*4.5. Curvature-affected viscosity: lipid packing effects*

POPC, a commonly used and physiologically representative phospholipid, restricts water from penetrating through its hydrophilic head group to its hydrophobic tails through self-assembly with 0.6 nm$^2$/lipid. In a POPC bilayer, the area of the headgroups and the volume of the tails results in a model membrane with no intrinsic curvature. The POPC molecular shape can be approximated as a cylinder (Figure 2). If a monolayer of POPC is forced to bend with a negative curvature, there is a greater packing of the head groups and extra volume for the tails. Alternatively, if a monolayer of POPC is forced to bend with a positive curvature, the acyl tails of POPC are compressed to a smaller volume, and the lipid head groups are stretched over a larger area. In both cases, parts of the POPC molecules are crowded more tightly than they would otherwise prefer, and a restriction to the free diffusion of lipids through the monolayer may result. No experiments to date have been able to distinguish the effects of positive curvature *versus* negative curvature on the effective membrane viscosity. However, net effects of inducing a membrane bud with both negative and positive curvature seem to be a net increasing the local membrane viscosity and slowing of the lipid mobility [18].

## 5. Conclusions

Membrane bending at physiological length scales provides numerous challenges for experimental observation. Diverse super-resolution optical techniques are providing resolution to the features of nanoscale membrane topography; however, the dynamical effects of curvature remain largely unknown. The ability for super-resolution optical techniques such as polarized localization microscopy to reveal nanoscale membrane bending is expanding the experimental capabilities for membrane curvature detection. The capacity to engineer membrane bending through the creation of supported lipid bilayers draped over lithographically defined substrates and polystyrene nanoparticles, allow for the experimental creation of membrane topographies that are analogous to endocytosis and exocytosis. Through the use of SPT, the spatial resolution of lipid mobility and membrane bending can be resolved with higher precision than detectable with FCS or FRAP. In particular, the fitting of the histogram of single-step sizes distribution enables the calculation of the lipid diffusion coefficients that are corrected for the localization uncertainty and the camera blur. Future experimental implementations of PLM and SPT will reveal the effects of membrane bending on the membrane viscosity and lipid mobility. Through asymmetric model membranes, the specific contribution of each bilayer leaflet will be determined in the nanoscale budding membrane. The sum of these results will contribute to the greater understanding of membrane biophysics and the mechanisms of cellular regulation of membrane topography.

**Acknowledgments:** The authors thank Zhi Mei for support with SEM imaging. A.M.K. was funded by Thomas C. Rumble Fellowship Award. Financial support was provided by Wayne State University laboratory startup funds, Richard J. Barber, and a CAREER award from the National Science Foundation (DMR1652316).